\documentclass[12pt,preprint]{aastex}

\slugcomment{To appear in ApJ}

\shorttitle{Fullerene-containing MCPNe}
\shortauthors{Garc\'{\i}a-Hern\'andez et al.}

\begin{document}


\title{The formation of fullerenes: clues from new C$_{60}$, C$_{70}$, and 
(possible) planar C$_{24}$ detections in Magellanic Cloud Planetary Nebulae}


\author{D. A. Garc\'{\i}a-Hern\'andez\altaffilmark{1,2}, S.
Iglesias-Groth\altaffilmark{1,2}, J. A. Acosta-Pulido\altaffilmark{1,2}, A.
Manchado\altaffilmark{1,2,3}, P. Garc\'{\i}a-Lario\altaffilmark{4}, L.
Stanghellini\altaffilmark{5}, E. Villaver\altaffilmark{6}, Richard A.
Shaw\altaffilmark{5} and F. Cataldo\altaffilmark{7,8}}


\altaffiltext{1}{Instituto de Astrof\'{\i}sica de Canarias, C/ Via L\'actea
s/n, 38200 La Laguna, Spain; agarcia@iac.es, amt@iac.es}
\altaffiltext{2}{Departamento de Astrof\'{\i}sica, Universidad de La Laguna (ULL), E-38205 La Laguna, Spain}
\altaffiltext{3}{Consejo Superior de Investigaciones Cient\'{\i}ficas, Spain}
\altaffiltext{4}{Herschel Science Centre. European Space Astronomy Centre,
Research and Scientific Support Department of ESA. Villafranca del Castillo,
P.O. Box 50727. E-28080 Madrid. Spain; Pedro.Garcia-Lario@sciops.esa.int }
\altaffiltext{5}{National Optical Astronomy Observatory, 950 North Cherry
Avenue, Tucson, AZ 85719, USA; shaw@noao.edu, letizia@noao.edu}
\altaffiltext{6}{Departamento de F\'{\i}sica Te\'orica C-XI, Universidad
Aut\'onoma de Madrid, E-28049 Madrid, Spain; eva.villaver@uam.es}
\altaffiltext{7}{Istituto Nazionale di Astrofisica - Osservatorio Astrofisico di
Catania, Via S. Sofia 78, 95123 Catania, Italy; franco.cataldo@fastwebnet.it}
\altaffiltext{8}{Actinium Chemical Research, Via Casilina 1626/A, 00133 Rome,
Italy}


\begin{abstract}
We present ten new Spitzer detections of fullerenes in Magellanic Cloud
Planetary Nebulae, including the first extragalactic detections of the C$_{70}$
molecule. These new fullerene detections together with the most recent
laboratory data permit us to report an accurate determination of the C$_{60}$
and C$_{70}$ abundances in space. Also, we report evidence for the possible
detection of planar C$_{24}$ in some of our fullerene sources, as indicated by
the detection of very unusual emission features coincident with the strongest
transitions of this molecule at $\sim$6.6, 9.8, and 20 $\mu$m. The infrared
spectra display a complex mix of aliphatic and aromatic species such as
hydrogenated amorphous carbon grains (HACs), PAH clusters, fullerenes, and small
dehydrogenated carbon clusters (possible planar C$_{24}$). The coexistence of such a
variety of molecular species supports the idea that fullerenes are formed from
the decomposition of HACs. We propose that fullerenes are formed from the
destruction of HACs, possibly as a consequence of shocks driven by the fast
stellar winds, which can sometimes be very strong in transition sources and young
PNe. This is supported by the fact that many of our fullerene-detected PNe show
altered [NeIII]/[NeII] ratios suggestive of shocks as well as P-Cygni profiles
in their UV lines indicative of recently enhanced mass loss.
\end{abstract}





\section{Introduction}

Fullerenes had been predicted since the 70s (Osawa et al. 1970) as a possible
carbon assemblage in the Universe, but have remained elusive in the
astrophysical context until recently, when C$_{60}$ and C$_{70}$ were detected
in the young Planetary Nebula (PN) Tc 1 (Cami et al. 2010) under conditions
which were claimed to be in agreement with the original laboratory studies on
the formation of fullerenes (Kroto et al. 1985; Kratschmer et al. 1990; de Vries
et al. 1993), this is under hydrogen-poor conditions. More recently,
Garc\'{\i}a-Hern\'andez et al. (2010, 2011) have challenged our understanding of
the fullerene formation in space, showing that, contrary to general expectation,
fullerenes (usually accompanied by polycyclic aromatic hydrocarbon (PAH)
emission) are efficiently formed in H-rich circumstellar environments. The
suggestion was made that both fullerenes and PAHs can be formed by the
photochemical processing of hydrogenated amorphous carbon grains (HACs), which
should be a major constituent in the circumstellar envelope of C-rich evolved
stars. The detection of fullerenes in a larger sample of PNe is essential to
answer key questions about fullerene formation in space. In this context, we
must note that C$_{60}$ has also been detected in the interstellar medium
(Sellgren et al. 2010), and in a proto-PN (Zhang \& Kwok 2011). The latter
raises the exciting possibility that considerable gas-phase physical and
chemical processing of dust occurs just after the Asymptotic Giant Branch (AGB)
phase, during which most of the circumstellar dust is formed (e.g., Wallerstein
\& Knapp 1998). Fullerenes, being relatively hardy molecules (Cataldo et al.
2009), may survive indefinitely in space playing an important role in
circumstellar/interstellar chemistry.

From the experimental point of view, it is important to know the temperature
dependence of the wavelength (and width) and of the strength (e.g., molar
absorptivity) of the C$_{60}$ and C$_{70}$ infrared bands. Very recent
laboratory spectroscopy of C$_{60}$ and C$_{70}$ (Iglesias-Groth et al. 2011)
has provided this detailed information. Moreover, PNe in the Magellanic Clouds
(MCPNe hereafter) offer the opportunity of obtaining an accurate determination
of the fullerene abundances in space. In this letter, we report ten new Spitzer
detections of C$_{60}$ and C$_{70}$ fullerenes in MCPNe together with the first
possible detection of planar C$_{24}$ in space, which provide us with crucial
information about fullerene formation in PNe. 

\section{Infrared spectra of fullerene-containing MCPNe}

Garc\'{\i}a-Hern\'andez et al. (2010) reported that SMP-SMC 16 showed obvious
C$_{60}$ bands in its Spitzer/IRS spectrum, acquired by Stanghellini et al.
(2007) within a sample of 41 MCPNe. Here we have re-analized in more detail the
full sample of IRS spectra mentioned above and those from an additional sample
of 25 MCPNe also observed with Spitzer/IRS (Bernard-Salas et al. 2009). We have
obtained the corresponding residual spectra by subtracting the dust continuum
emission between 5 and 38 $\mu$m, which is represented by 3-7 order polynomials
fitted at spectral locations free from any dust or gas feature. A careful
inspection of the residual spectra reveals the fullerene signatures in 10
additional objects (6 SMC and 4 LMC PNe, see Table 1). Flux measurements
(errors $\leq$30$-$40\%) were done in the residual spectra by
subtracting a local baseline and fitting gaussians (Garc\'{\i}a-Hern\'andez et
al., in preparation; hereafter, GH2011). We list our targets in Table 1, where
we also provide the central star temperatures, the H$_{\beta}$ fluxes, carbon
abundances, electron temperatures and densities, total hydrogen and carbon
masses of the nebulae. Most of these newly detected C$_{60}$-containing MCPNe
also show broad dust emission features at $\sim$11.5 and 30 $\mu$m, which
sometimes are accompanied by other broad dust features at 6-9 and 15-20 $\mu$m
(see Stanghellini et al. 2007; Bernard-Salas et al. 2009). Bernard-Salas et al.
(2009) found that the strength of the 11.5 $\mu$m feature decreases with
increasing [NeIII]/[NeII]\footnote{The ratio of the flux from [Ne III] 
$\lambda15.5~\mu$m to that of [Ne II] $\lambda12.8~\mu$m (hereafter, [Ne
III]/[Ne II]).} ratio (taken as an indicator of the radiation field hardness).
However, SMC 16 and LMC 02 display the lowest [NeIII]/[NeII] ratios (Table 2) in
our sample and the 11.5$\mu$m feature is barely detected in SMC 16 and
completely absent in LMC 02. A low [NeIII]/[NeII] ratio may be explained if the
[NeII] 12.8$\mu$m line is excited by shocks (see Sect. 3). 

In Fig. 1 we display the residual spectra, where the expected positions of the
four C$_{60}$ bands are marked with dashed vertical lines. A few of the C$_{60}$
sources also display relatively strong and isolated C$_{70}$ features at 12.6,
13.8, 14.9, 15.6, 17.3, 17.6, and 18.7 $\mu$m. In particular, C$_{70}$ is
clearly detected in SMC 24 and LMC 02 for which high-resolution (R$\sim$600)
Spitzer spectra are available (Fig. 2). We can neither confirm nor exclude the
presence of C$_{70}$ in other sources because of the much lower resolution of
their Spitzer spectra, similarly to what we discovered for SMC 16
(Garc\'{\i}a-Hern\'andez et al. 2010). SMC 24 and LMC 02 display an infrared
spectrum richer than Tc 1 (Fig.2) with several unidentified features such as
those at 16.0, 16.1, 16.6, 18.0, 18.3, and 18.4 $\mu$m. The identification of
these features will be studied elsewhere

The Spitzer spectra of fullerene MCPNe show that a complex mixture of aliphatic
and aromatic species is usually present. A broad 6$-$9 $\mu$m emission, which
can be attributed to HACs, large PAH clusters or very small grains (Tielens
2008), is seen with different strengths in most of our sources (Fig. 1). The
6$-$9 $\mu$m feature is more intense in SMC 16, 24, and LMC
02\footnote{Bernard-Salas et al. (2009) identified the 6$-$9 $\mu$m emission in
SMC 24 and LMC 02 with HACs.} and displays a shape different to the other
sources. Narrower features at $\sim$6.3 and 7.8$-$7.9 $\mu$m are also detected
(e.g., SMC 20, 13, LMC 25, 48, 99). The latter features are seen in a few
Galactic post-AGB stars and can be attributed to large PAHs or to hydrocarbons
(as a relatively unprocessed mixture of aliphatics and aromatics; Sloan et al.
2007). Additional weak features at 6.9 and 7.3 $\mu$m that have been identified
with aliphatic hydrocarbons and/or HACs (Sloan et al. 2007 and references
therein) are also detected in SMC 24. 

Interestingly, some of our sources (SMC 13, 16, 24, LMC 48, 02) display an
unidentified infrared feature (UIR) at $\sim$6.6 $\mu$m. This feature is
accompanied by other unidentified features at $\sim$9.8 and 20 $\mu$m in SMC 24
and LMC 02 (see Fig. 1). The wavelength position of the three UIR features seen
in SMC 24 and LMC 02 agree very well with the strongest transitions of planar
C$_{24}$ as recently reported by Kuzmin \& Duley (2011) from calculated infrared
spectra using first principles density functional techniques. In SMC 24, the
observed flux ratios are F(6.6)/F(9.8)=3.3 and F(6.6)/F(20.0)=1.5, which compare
well with the theoretical integrated intensity (km/mole) ratios of 3.8 and 2,
respectively (Kuzmin \& Duley 2011). However, the 6.6 $\mu$m feature is not
clearly distinguised from the 6$-$9$\mu$m emission in LMC 02 and the observed
9.8 $\mu$m feature is $\sim$3 times more intense than that at 20 $\mu$m. As in
the case of the C$_{60}$ and C$_{70}$ fullerenes (Iglesias-Groth et al. 2011),
one expects the intensity, width, and position of the infrared features of
planar C$_{24}$ to be temperature dependent. Planar C$_{24}$ is a very stable
molecule (more than the C$_{24}$ fullerene although Jones \& Seyfert (1997)
show that other C$_{24}$ conformations may be more stable than planar C$_{24}$).
The detection of these very unusual $\sim$6.6, 9.8, and 20 $\mu$m features in
fullerene sources represents the first possible detection of planar C$_{24}$ in
space. However, this can only be confirmed through laboratory infrared
spectroscopy at different temperatures, which is extremely difficult because of
the high reactivity of C$_{24}$.

\section{CLOUDY simulations}

Recent laboratory spectroscopy of C$_{60}$ and C$_{70}$ fullerenes show that
C$_{70}$ displays a strong transition at $\sim$7.0 $\mu$m (Iglesias-Groth et al.
2011). The 7.0$\mu$m C$_{70}$ transition is one of the strongest (with a
strength very similar to the other $\sim$15.6 and 18.7$\mu$m C$_{70}$
transitions). The 7.0$\mu$m C$_{60}$ band is, however, less intense (a factor of
$\sim$3) than that at 18.9 $\mu$m. Indeed, the 7.0$\mu$m line intensity observed
in fullerene PNe cannot be explained by C$_{60}$ alone and it has been assumed
to be strongly contaminated by [ArII]6.99$\mu$m (Cami et al. 2010;
Garc\'{\i}a-Hern\'andez et al. 2010). We show here that the 7.0$\mu$m emission
seen in our sample can not be assigned to [ArII] only, rather, it should be 
attributed to a combination of C$_{60}$ and C$_{70}$, in agreement with the
recent laboratory data. To this end, we have used a photoionization code to
model the relative intensity of the mid-infrared lines observed in the
Spitzer/IRS spectra. Calculations were performed with CLOUDY (Ferland et al.
1998). We do not intend to find the best solution for each of the observed PNe
but to make general predictions using the models. As ionizing continuum we
assumed blackbody and synthetic stellar spectra (Rauch 2003) with temperatures
from 25,000 to 50,000 K. Typical PN values for the bolometric luminosity
(10$^{4}$ L$_{\odot}$) and for the radius of the photoionized region (0.1 pc)
are adopted (Mendez et al. 1992; Villaver et al. 2002, 2007). There is no way to
explain the extremely low [NeIII]/[NeII] ratios observed in some of our sources
(see below), even taking into account large variations of these parameters
(GH2011). We also assumed a density of 1800 cm$^{-3}$ at r=0.1 pc, which
varies as r$^{-2}$  along the nebula and the chemical abundances by Shaw et
al.(2010).

Our model predictions are compared with the observed line intensity ratios
([ArIII]/[HI] vs. [ArII/HI] and [SIV]/[SIII] vs. [NeIII]/[NeII]) in Fig. 3. The
7.0$\mu$m feature cannot be explained by [ArII]6.99$\mu$m emission (even taking
into account maximum errors of 40\% in the observed line flux ratios) and the
observed 7.0$\mu$m emission is much higher than expected from photoionization
from the central stars (T$_{eff}$$>$31,000 K; Table 1). Assuming that the
7.0$\mu$m emission is due to the Ar lines, the Ar abundances for these PNe would
be unrealistically high. Another important conclusion can be extracted from the
plot [SIV]/[SIII] vs. [NeIII]/[NeII] (Fig.3). The low [NeIII]/[NeII]
ratios\footnote{Note that the [SIV]/[SIII] ratio in LMC 02 is much lower than
expected for its T$_{eff}$ because the observed 18.7$\mu$m emission is dominated
by C$_{70}$.} observed in many of our PNe cannot be explained by
photoionization. This discrepancy can be explained if the [NeII] 12.8$\mu$m line
is shock-excited by the fast stellar winds from the central stars (e.g.,
Hartigan et al. 1987; Molinari \& Noriega-Crespo 2002; van Boekel et al. 2009).
A [NeII] shock-excited origin is also suggested from the common detection of
P-Cygni profiles in the UV lines of our PNe (see Sect.5); a clear indication of
recently enhanced mass loss and thus, of possible shocks. 

\section{Excitation mechanism and abundances}

Previous works suggest that the fullerene emission in PNe does not originate
from free gas-phase fullerene molecules, but from molecular carriers attached to
solid material (Cami et al. 2010; Garc\'{\i}a-Hern\'andez et al. 2010). However,
the fullerene emission seen in sources with cooler stars (T$_{eff}$$<$25,000 K)
such as reflection nebulae (Sellgren et al. 2010), RCB stars
(Garc\'{\i}a-Hern\'andez et al. 2011), and proto-PNe (Zhang \& Kwok 2011) may be
explained by UV-excited C$_{60}$ molecules in the gas phase. Sellgren et al.'s
(2010) calculations indicate that the intensities of the C$_{60}$ 7.0 and 8.5
$\mu$m bands versus the 18.9 $\mu$m band are very sensitive to the energy of the
absorbed UV photons; with the C$_{60}$ F(17.4)/F(18.9) ratio being not sensitive
to the absorbed photon energy. Assuming a solid-state origin for the observed
C$_{60}$ bands instead, we have determined the fluxes that are actually due to
the 7.0 and 8.5 $\mu$m C$_{60}$ transitions\footnote{The observed 8.5$\mu$m
fluxes cannot be completely explained by C$_{60}$ in four PNe (SMC 13, 24, LMC
02, 99).} from the corresponding excitation diagrams (see below). The C$_{60}$
F(7.0)/F(18.9) and F(8.5)/F(18.9) ratios are listed in Table 2 for a proper
comparison with Sellgren et al.'s (2010) predictions. Interestingly, the
C$_{60}$ F(7.0)/F(18.9) and F(8.5)/F(18.9) ratios indicate typical photon
energies of $\leq$5 eV for all sources in our sample independently of the
effective temperature ($>$31,000 K, when available) of the central star (see
Sect.5). 

We determine the fullerene temperatures for each PN from the flux measurements
of the C$_{60}$ and C$_{70}$ transitions and adopting the temperature dependence
of the absorptivity for each C$_{60}$ and C$_{70}$ transition (Iglesias-Groth et
al. 2011). We developed an iterative process to find the best correlation
coefficient of the linear relation in the Boltzmann excitation diagram, from
which the temperature and the total number of emitting fullerene molecules can
be estimated (GH2011). The initial temperatures were assumed in the range
100$-$1000 K and a few iterations were required for a quick convergence. The
four mid-IR transitions of C$_{60}$ were considered, allowing the fraction of
the flux of the 7$\mu$m C$_{60}$ transition as a free parameter. In the case of
C$_{70}$, we considered the 3 or 4 transitions less contaminated by other
species, obtaining also the fraction of the 7$\mu$m C$_{70}$ flux. We found very
good correlation coefficients (0.89$-$0.99) for the C$_{60}$ and C$_{70}$
fullerenes. The temperatures and the C$_{60}$ and C$_{70}$ contributions to the
observed 7$\mu$m flux are listed in Table 2. The derived fullerene excitation
temperatures are in the range $\sim$200$-$600 K and the emission may come from
solid-state fullerenes (e.g., molecules attached to dust grains). Our excitation
temperatures indicate that C$_{70}$ is cooler than C$_{60}$. 

The fullerene-sources in the MCs provide a reliable estimate of the C$_{60}$ and
C$_{70}$ abundances in PNe because the distances to the SMC (61 kpc; Hilditch
2005) and LMC (48.5 kpc; Freedman \& Madore 2010) are accurately known, and
because reliable C atomic abundances from UV spectra are available for  many of
these sources in the literature (Table 1). By using the above distances and
following Garc\'{\i}a-Hern\'andez et al. (2010)\footnote{The H masses are
estimated from the available electronic temperatures and densities and the
observed H$_{\beta}$ fluxes (Table 1).}, the C$_{60}$/C and C$_{70}$/C
abundances were estimated. We find  C$_{60}$/C$\sim$0.003$-$0.29 (average of
$\sim$0.07) and C$_{70}$/C$\sim$0.001$-$0.07 (average of $\sim$0.03). Our
average C$_{60}$/C$\sim$0.07 is consistent with the C$_{60}$ abundance recently
derived in a proto-PN (Zhang \& Kwok 2011). By assuming a distance of 2 kpc for
Tc 1 (as Cami et al. 2010), we estimate C$_{60}$/C=0.04 and C$_{70}$/C=0.005.

\section{Formation of fullerenes}

Laboratory studies show that the C$_{70}$/C$_{60}$ ratio can tell us whether the
formation of fullerenes is ``bottom-up" or ``top-down", meaning growth from
smaller to larger, or decay from larger to smaller units, respectively (e.g.,
Mansurov 2011). In combustion, one can get a high ratio (0.26$-$8.8), while in
carbon vapor condensation the ratio is usually low (0.02$-$0.18) (e.g., Mansurov
2011). In addition, vapor carbon condensation experiments under H-poor
conditions indicate that C$_{60}$ fullerenes are more than 50\% of total large
carbon clusters (Kroto et al. 1985). The low C$_{70}$/C$_{60}$ of
$\sim$0.02$-$0.2 (Table 2) seems to indicate that fullerenes are formed from
vapor carbon condensation. However, the fact that all known fullerene-containing
PNe are H-rich and that C$_{60}$ only represents a very small percentage of the
total carbon in these stars is against the idea that fullerenes are formed under
H-poor conditions. Garc\'{\i}a-Hern\'andez et al. (2010, 2011) have suggested
that fullerenes can be formed from the decomposition of HACs in order to explain
the simultaneous detection of C$_{60}$ and PAHs in H-rich circumstellar
envelopes. Unfortunately, the C$_{70}$/C$_{60}$ ratio cannot be measured in the
laboratory when fullerenes are produced from laser vaporization of HACs films.
This is because it is common to see variations in the relative intensities of
the C$_{60}$ and C$_{70}$ mass peaks even in mass spectra recorded on the same
sample and under the same experimental conditions (W. Duley 2011, private
communication). 

Laboratory experiments show that the decomposition of HACs is sequential with
small dehydrogenated PAH molecules released first, and then fullerenes and large
PAH clusters (Scott et al. 1997). We have shown here that fullerene MCPNe show a
complex mixture of aliphatic and aromatic species such as HACs, PAH clusters,
fullerenes, and small dehydrogenated carbon clusters (possible planar C$_{24}$).
The coexistence of this large variety of molecular species in PNe with
fullerenes is strongly supported by the laboratory results obtained by Scott and
colleagues, indicating that fullerenes (and PAHs) are formed from the
vaporization of HACs.

However, the excitation mechanism of the fullerene emission is still not clear.
Neutral fullerenes in MCPNe seem to be not excited by the bulk of UV photons
coming from the central star (T$_{eff}$$>$31,000 K, when available); the photon
energies involved to explain the observed C$_{60}$ intensity ratios (Table 2)
should be $\leq$5 eV according to Sellgren et al. (2010). Our results are not
consistent with Sellgren et al. (2010) who concluded that the fullerene emission
observed in the reflection nebula NGC 7023 is due to gas-phase C$_{60}$
molecules excited by UV photons of $\sim$10 eV, in agreement with the bulk of UV
photons coming from the exciting star with T$_{eff}$=21,000 K. An alternative
explanation may be that fullerenes are collisionally excited during the HACs
vaporization process (i.e., grain-grain collisions) and emit even when they are
still attached to the HACs surface. Other possibilities are the dehydrogenation
of the molecules present in the HACs surface, before they are released to the
gas phase or even shock-excited gas-phase fullerenes\footnote{Possibly this
is also the case for the planar C$_{24}$ emission observed.}.Presumably
shocks are not energetic enough to ionize the fullerenes, but this is an open
question and the shock-excited hypothesis needs further more detailed modeling
work. The lack of mid-IR C$_{60}$$^{+}$ features in fullerene-containing
sources, may imply that fullerenes are either quickly destroyed in the gas phase
or never leave the grains as free molecules. This ambiguity arises because
fullerenes may be created in the vaporization process, rather than being
pre-existing molecules in the original HACs sample. Indeed, the laboratory
experiments use a laser at 4 eV to simulate HACs vaporization by interstellar
shocks. Our Fig. 3. show that the observed [NeIII]/[NeII] ratios in many of the
circumstellar shells of fullerene-containing PNe cannot be explained by
photoionization and they may be strongly altered by shocks. Five
fullerene-containing sources (SMC 15, 16, 20, LMC 25, 48) display P-cygni
profiles in their UV lines, which are indicative of on-going mass loss with
high-velocity ($>$1000 kms$^{-1}$) strong winds, as confirmed by HST UV
spectroscopy (Stanghellini et al. 2005, 2009). Furthermore, given the low S/N in
the HST ACS spectra, the presence of P-cygni profiles in SMC 24 and SMC 18 is
hinted although could not be confirmed. LMC 99 and LMC 02 lack the necessary
diagnostics to determine the presence of P-Cygni profiles in the UV. SMC 13,
which does not show any P-Cygni profiles, displays the lowest C$_{60}$/C
abundance, suggesting that fullerene formation is less efficient. On the other
hand, fullerene production seems to be more efficient in SMC 16 and LMC
02\footnote{SMC 24 with intense 6$-$9$\mu$m emission also displays a
[NeIII]/[NeII] ratio lower than expected for its T$_{eff}$ of 37,800 K} with the
stronger outflows (as indicated by their extremely low [NeIII]/[NeII] ratios)
and this is the subset of sources which display as well  the more intense 6$-$9
$\mu$m emission that may be attributed to HACs or PAH clusters.

In summary, we propose that shocks (i.e., grain-grain collisions) driven by the
strong stellar winds are triggering the HACs processing and that fullerenes (as
well as other complex aromatic and aliphatic species like possibly planar
C$_{24}$ molecules that may have been detected for the first time) evolved from
the vaporization of HACs. Our interpretation can be observationally tested
through high-resolution (R$\sim$30,000) observations around [Ne II]12.8$\mu$m,
which will confirm/refute the possible [Ne II] shock-excited origin in
fullerene-containing PNe.



\acknowledgments

We thank the anonymous referee and W. Duley for suggestions that help to improve the
paper. 




{\it Facilities:} \facility{Spitzer:IRS}.

\clearpage

\begin{deluxetable}{ccccccccc}
\tabletypesize{\scriptsize}
\tablecaption{The sample of fullerene-containing MCPNe.\label{tbl-2}}
\tablewidth{0pt}
\tablehead{
\colhead{Object} &  \colhead{T$_{eff}$} &  \colhead{log(F(H$\beta$))} & \colhead{A(C)} 

&  \colhead{T$_{e}$} & \colhead{N$_{e}$} & \colhead{Ref.$^{a}$} &  \colhead{M$_{H}$} & \colhead{M$_{C}$}   \\
\colhead{} &  \colhead{(K)}  & \colhead{} & \colhead{log(C/H)+12} &
\colhead{(10$^{4}$ K)} &  \colhead{(cm$^{-3}$)} 
& \colhead{} & \colhead{(M$_{\odot}$)} & \colhead{(10$^{-3}$ M$_{\odot}$)}
}
\startdata
SMC 13       & 31,300 &  -12.59 &  8.73 &  1.28  & 2900    &  (1,2,3,4,4)        &  0.702  & 4.520 \\
SMC 15       & $\dots$&  -12.45 &  8.26 &  1.41  & 7500    &  ($\dots$,5,3,6,7)	 &  0.270  & 0.588 \\
SMC 16       & $\dots$&  -12.74 &  8.19 &  1.18  & 4400    &  ($\dots$,5,3,6,7)	 &  0.210  & 0.392 \\
SMC 18       & $\dots$&  -12.66 &  8.31 &  1.18  & 3600    &  ($\dots$,2,3,4,4)	 &  0.381  & 0.934 \\
SMC 20       & $\dots$&  -12.47 &  8.25 &  1.38  & 3900    &  ($\dots$,2,3,4,4)	 &  0.497  & 1.061 \\
SMC 24$^{c}$ & 37,800 &  -12.66 &  8.18 &  1.16  & 2800    &  (1,2,3,4,4)	 &  0.411  & 0.747 \\
SMC 27       & 43,300 &  -12.51 &$\dots$&  1.27  & 3650    &  (1,2,$\dots$,4,4)	 &  0.471  & $\dots$ \\
LMC 02$^{c}$ & 39,000 &  -13.18 &  8.14 &  1.21  & 5000    &  (8,7,9,6,8)	 &  0.048  & 0.079 \\
LMC 25       & 33,700 &  -12.39 &  8.29 &  1.56  & 3300    &  (10,11,12,13,13)	 &  0.626  & 1.465 \\ 
LMC 48       & $\dots$&  -12.48 &  8.40 &  1.00$^{b}$ & 1900 & ($\dots$,5,12,$\dots$,6)  &  0.844  & 2.540 \\
LMC 99       & $\dots$&  -12.54 &  8.77 &  1.22  & 3600    & ($\dots$,14,15,6,6) &  0.434  & 3.040 \\
\hline
Tc 1$^{d}$   & 34,100 &  -10.73 &  8.59 & 0.90   & 3000    &  (16,17,18,18,18)   &  0.0535  & 0.250  \\
\enddata
\tablenotetext{a}{References for T$_{eff}$, log(F(H$\beta$)), C abundance, T$_{e}$, and N$_{e}$.}
\tablenotetext{b}{T$_{e}$ for LMC 48 is assumed to be 10$^{4}$ K.}
\tablenotetext{c}{High-resolution Spitzer/IRS spectra were obtained and it was possible to detect C$_{70}$ unambiguously.}
\tablerefs{(1) Villaver et al. (2004); (2) Stanghellini et al. (2003); (3) Stanghellini et al. (2009); (4) Shaw et al. (2010); (5) Shaw et al. (2006);
(6) Leisy \& Dennefeld (2006); (7) Dopita et al. (1988); (8) Herald \& Bianchi (2004); (9) Milanova \& Kholtygin (2009); (10) Villaver et al. 2003;
(11). Stanghellini et al. (2002); (12) Stanghellini et al. (2005); (13) Shaw et al. (2011, in prep.); (14) R. Shaw (2011, priv. comm.); (15) Leisy \&
Dennefeld (1996); (16) Preite-Mart\'{\i}nez et al.(1989; (17) Cahn et al. (1992); (18) Williams et al. (2008).}
\end{deluxetable}

 \clearpage

\begin{deluxetable}{cccccccccccc}
\tabletypesize{\scriptsize}
\setlength{\tabcolsep}{0.02in} 
\tablecaption{C$_{60}$ intensity ratios and fullerene abundances and temperatures.\label{tbl-2}}
\tablewidth{0pt}
\tablehead{ \colhead{Object} &  
\colhead{F(7.0)/F(18.9)$^{a}$} &
\colhead{F(8.5)/F(18.9)$^{a}$} &
\colhead{F(17.4)/F(18.9)$^{a}$} & \colhead{[NeIII]/[NeII]} &
\colhead{C$_{60}$/C} &  \colhead{f$_{C60}$$^{b}$} & 
\colhead{T$_{C60}$} & \colhead{C$_{70}$/C}&  \colhead{f$_{C70}$$^{b}$} & \colhead{T$_{C70}$} & \colhead{C$_{70}$/C$_{60}$}\\

\colhead{} & \colhead{} &\colhead{} &\colhead{} &\colhead{} &

\colhead{(\%)}& \colhead{(\%)} & \colhead{(K)} & \colhead{(\%)} & \colhead{(\%)}& \colhead{(K)} & \colhead{}
}
\startdata
SMC 13  &   0.16  &0.32&    0.21    & 2.89  & 0.003  & 15$\pm$10 & 453$\pm$50 &$\dots$	&$\dots$ & $\dots$ & $\dots$	 \\
SMC 15  &   0.20  &0.30&    0.30    & 0.49  & 0.07   & 15$\pm$10 & 290$\pm$20 &$\dots$	&$\dots$ & $\dots$ & $\dots$	 \\
SMC 16  &   0.24  &0.33&    0.45    & 0.21  & 0.12   & 30$\pm$10 & 414$\pm$50 &$\dots$	&$\dots$ & $\dots$ & $\dots$     \\
SMC 18  &   0.16  &0.52&    0.44    & 1.16  & 0.03   & 15$\pm$10 & 365$\pm$35 &$\dots$	&$\dots$ & $\dots$ & $\dots$	 \\
SMC 20  &   0.05  &0.13& $\dots$    & 0.41  & 0.03   & 10$\pm$10 & 228$\pm$30 &$\dots$	&$\dots$ & $\dots$ & $\dots$	 \\
SMC 24  &   0.47  &0.18&    0.53    & 3.33  & 0.07   & 60$\pm$20 & 514$\pm$30 & 0.001	& 20$\pm$10& 383$\pm$60 &  0.02       \\
SMC 27  &   0.58  &0.48&    0.50    & 9.69  &$\dots$ & 50$\pm$10 & 553$\pm$40 &$\dots$	&$\dots$ & $\dots$ & $\dots$	 \\
LMC 02  &   0.48  &0.37&    0.71    & 0.42  & 0.29   & 30$\pm$10 & 502$\pm$20 & 0.07	& 10$\pm$10& 324$\pm$10 &  0.21		 \\	 
LMC 25  &   0.08  &0.26&    0.69    &13.19  & 0.02   & 50$\pm$20 & 300$\pm$25 &$\dots$	&$\dots$ & $\dots$ & $\dots$	 \\
LMC 48  &   0.07  &0.22&    1.13    &11.46  & 0.01   & 20$\pm$10 & 270$\pm$20 &$\dots$	&$\dots$ & $\dots$ & $\dots$	 \\
LMC 99  &   0.30  &0.56&    0.72    &16.82  & 0.02   & 50$\pm$10 & 446$\pm$25 &$\dots$	&$\dots$ & $\dots$ & $\dots$	 \\
\hline
Tc 1    &   0.26  &0.36&    0.53    & 0.05  & 0.04   & 25$\pm$10 & 415$\pm$30 & 0.005   & 40$\pm$10& 314$\pm$30 & 0.10	 \\
\enddata
\tablenotetext{a}{The ratio of fluxes in the C$_{60}$ bands with the wavelenghts in brackets.}
\tablenotetext{b}{C$_{60}$ and C$_{70}$ contributions to the observed 7$\mu$m flux.}
\end{deluxetable}

\clearpage

\begin{figure}
\includegraphics[angle=0,scale=.65]{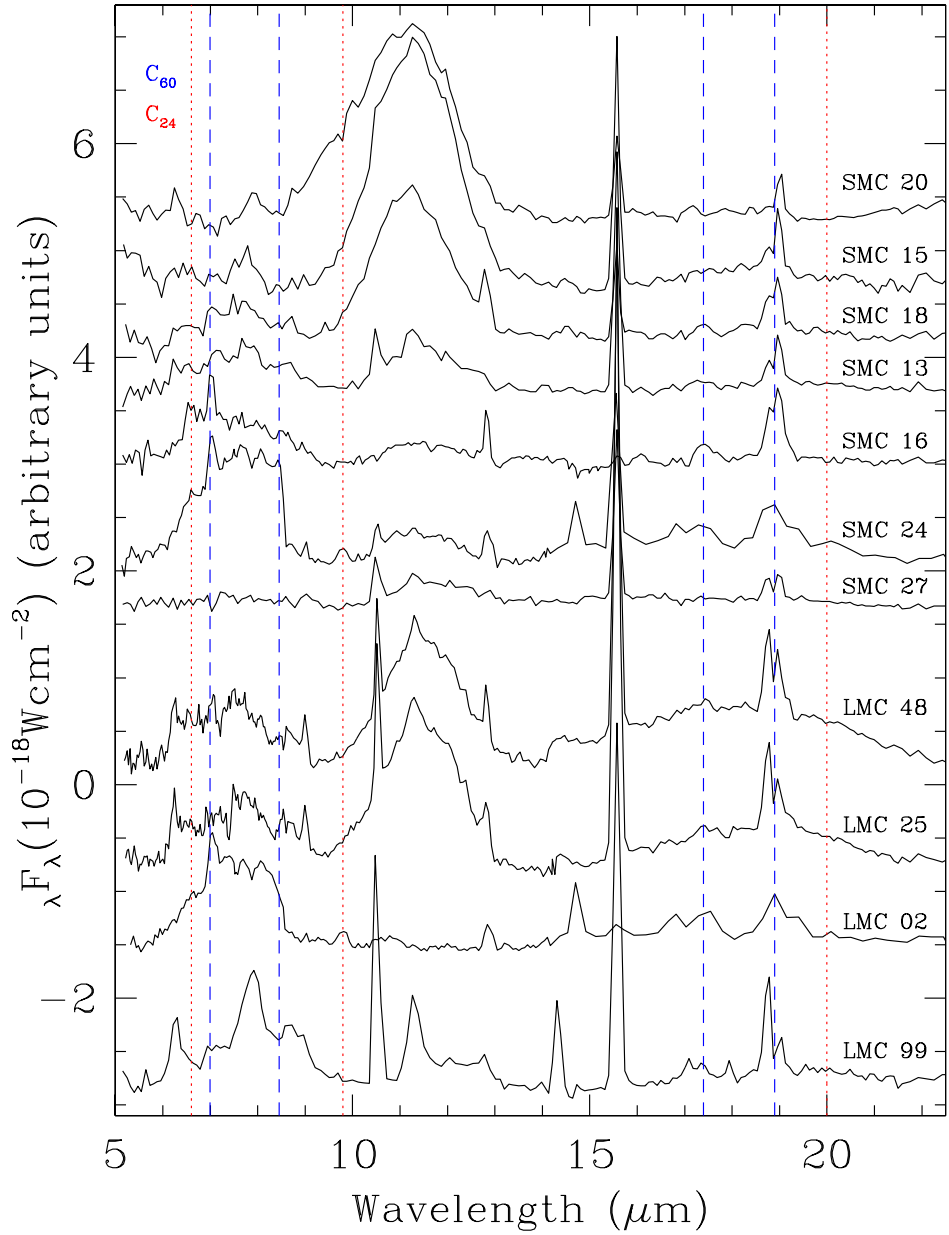}
\caption{Residual spectra ($\sim$5$-$23 $\mu$m) of fullerene-MCPNe. The band
positions of C$_{60}$ (dashed) and planar C$_{24}$ (dotted) are marked. The
spectra are displaced for clarity. \label{fig1}}
\end{figure}

\clearpage

\begin{figure}
\includegraphics[angle=0,scale=.65]{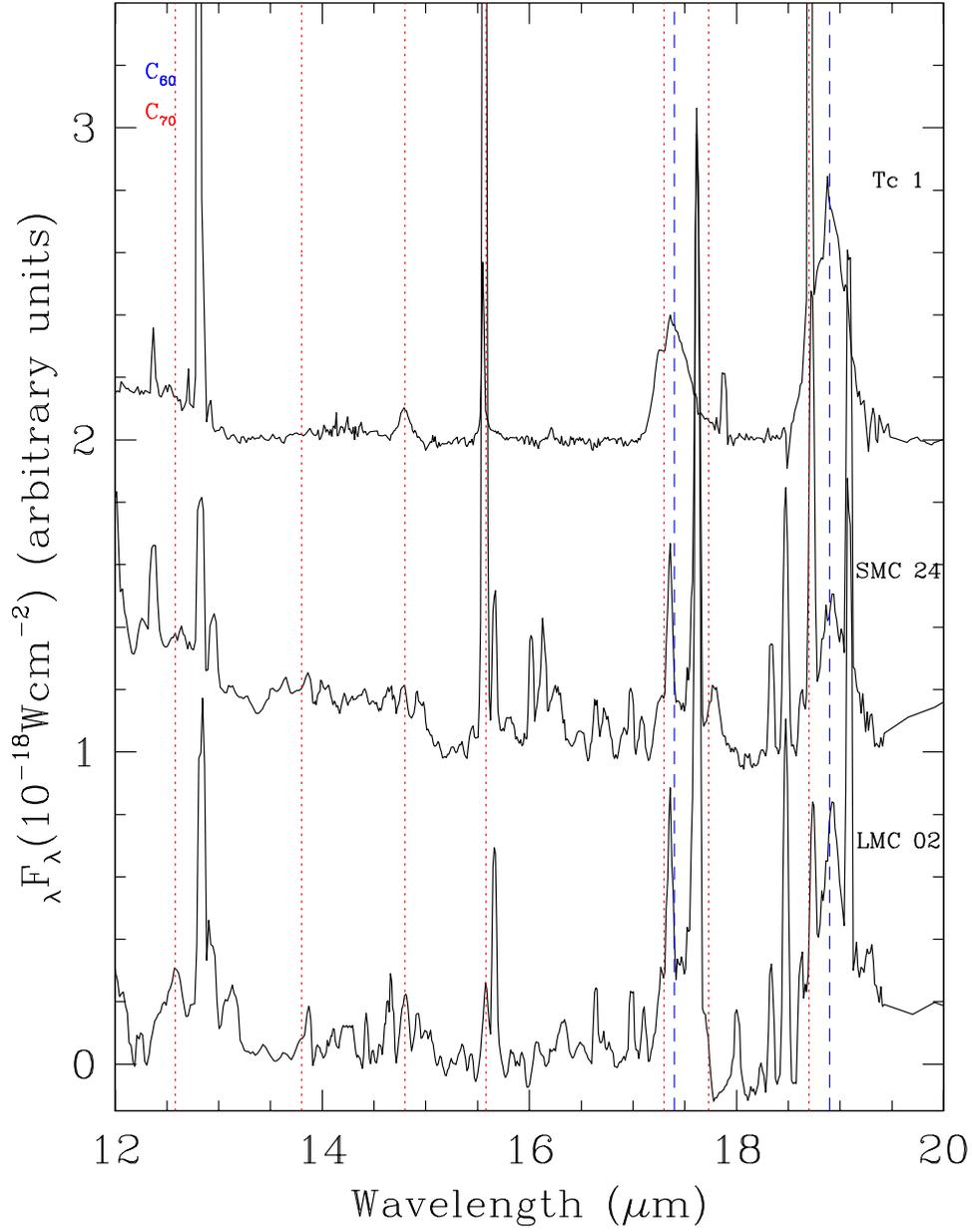}
\caption{Residual spectra ($\sim$10$-$23 $\mu$m) for C$_{70}$-MCPNe (in
comparison with Tc 1) where the C$_{60}$ (dashed) and C$_{70}$ (dotted) band
positions are marked. \label{fig2}}
\end{figure}

\clearpage

\begin{figure}
\includegraphics[angle=90,scale=.80]{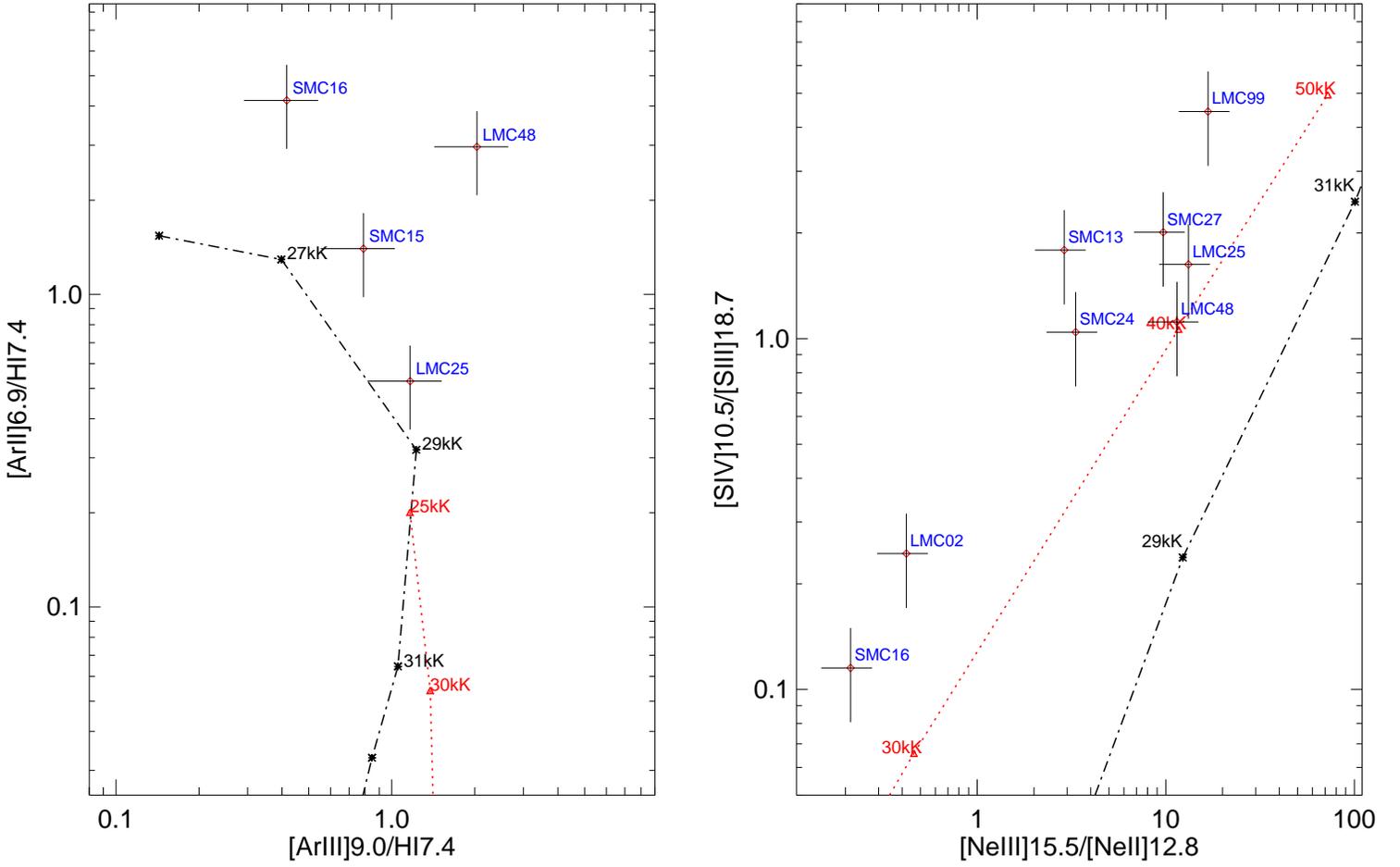}
\caption{CLOUDY predictions vs. observations. Predictions for different ionizing
continua such as blackbody spectra (red dotted line) and synthetic stellar
spectra (black dashed line) at several temperatures are shown (see text).
\label{fig3}}
\end{figure}

\end{document}